\documentclass[twocolumn,twoside]{article}
\usepackage{graphics}

\newcommand{\hb}{\\ \hspace*{2ex}}

\begin{document}
\title{THE RELATIVE WAVELENGTH INDEPENDENCE OF IR LAGS\\IN ACTIVE GALACTIC NUCLEI:
IMPLICATIONS FOR THE \\DISTRIBUTION OF THE HOT DUST}
\author{V.L.\,Oknyansky$^{1}$, C.M.\,Gaskell$^{2}$, E.V.\,Shimanovskaya$^{1}$\\[2mm] 
\begin{tabular}{l}
 $^1$ Moscow M.V. Lomonosov State University, Sternberg Astronomical Institute,\hb
 Moscow, 119991, Russian Federation, {\em oknyan@mail.ru}\\
 $^2$ Department of Astronomy and Astrophysics, University of California,\hb
 Santa Cruz, USA, {\em mgaskell@ucsc.edu}\\[2mm]
\end{tabular}
}
\date{}
\maketitle

ABSTRACT.
We show that, contrary to the simple prediction, most AGNs show at best only a small increase of lags in the $J$, $H$, $K$, and $L$ bands with increasing wavelength.  We suggest that a possible cause of this near simultaneity of the variability from the near-IR to the mid-IR is that the hot dust is in a hollow bi-conical outflow of which we only see the near side. Although most AGNs show near simultaneity of IR variability, there was at least one epoch when NGC~4151 showed the sharply increasing IR lag with the increase of the wavelength. This behaviour might also be present in GQ~Comae. We discuss these results briefly. The relative wavelength independence of IR lags simplifies the use of IR lags for estimating cosmological parameters.\\[1mm]
\\[2mm]

{\bf 1. Introduction}\\[1mm]

The variable near-IR radiation of active galactic nuclei (AGNs) is usually associated with the part of the optically-thick dusty torus closest to the central source (H\"onig \& Kishimoto 2011). The presence of such a torus is the key to explaining the observed differences in the spectra of type-1 and type-2 Seyfert nuclei by the torus blocking our direct view of the broad emission lines and the thermal continuum emitted by the accretion disc (AD).  It is also believed that the dusty torus radiates in the infrared, as a result of heating by shorter wavelength radiation from the accretion disc and X-ray emitting corona. Closer to the centre the dust is completely (or largely) sublimated and delayed  infrared variability gives us the estimate of radius of the ``dust holes'' around the central source (Oknyanskij \& Horne 2001), i.e., the radius of the region where the dust is absent. The work we present here is a continuation of our series of papers in which we measure the radius of the ``dust holes'' in the  NGC~4151 and other AGNs from the delay of the variability in the near infrared relative to the optical variability (see details and references at Oknyanskij et al. 1999, Oknyanskij \& Horne 2001, Oknyansky et al. 2014a,b). Despite the significant growth in theoretical and observational studies of AGNs in the IR, our knowledge of the dust, its origin, kinematics, and detailed morphology remains very incomplete.

The number of IR lags determined at different wavelengths for AGNs is not large. At the time of publications by Oknyansky et al. (1999), Oknyanskij \& Horne (2001), there were just a few estimates of IR lags at different wavelengths. The first results were somewhat controversial. Some objects (NGC~4151, GQ~Comae) showed a sharp increase of lag with wavelength in the IR. These differences in the lags (the lags in the $L$ band were 3 times longer than in the $K$ band) were in a good agreement with simple model predictions that cooler dust is farther from the centre than hotter dust (Barvainis 1987). Therefore Oknyansky \& Horne (2001) considered those results to be normal and used the observed increase to correct observed IR lags for red shift. However, at that time at least one object, Fairall~9, was known to have about the same lags for the $H$, $K$ and $L$ bands. Oknyanskij et al. (1999) interpreted this in the following way: if the minimum distance from the central source to the dust clouds corresponds to a brighter state of the nucleus well before the interval under study, then the differences in the lags for $K$ and $L$ will be insignificant, as was observed in the Fairall~9. If, alternatively, this minimum radius roughly corresponds to the maximum nuclear luminosity in the interval under consideration, then a marked difference in the lags for $K$ and $L$ must be observed, as it was the case for NGC~4151. This interpretation gives us observational prediction for NGC~4151 - at another time when it goes into a low state we will see similar IR lags for different IR wavelengths. This was indeed found for  NGC~4151 in subsequent papers by Oknyansky et al (2006, 2014a,b). In order to investigate the reliability of these results, we have been collecting all available time lag data in different IR bands for AGNs to see how often this phenomenon is observed.\\[2mm]

{\bf 2. Observed IR time lags}\\[1mm]

Table 1 gives published and new measurements of time lags between variations in different near- and mid-IR bands relative to variability of the optical or UV continuum for as many AGNs as possible. Some new results in Table \ref{tab1} were obtained applying our MCCF (Modified Cross-Correlation Function) method (Oknyanskij 1993)) to published photometric data. Examples of MCCFs are given just for NGC~7469 in Fig. 1. In most cases our results are in a good agreement with results obtained before using other methods. In a few cases, our new results (for Ark~120, Mrk~509) disagree with the values obtained before by Glass (2004). This is in part because of misprints in Glass (2004). We have also used some unpublished optical data.  $JHKL$ photometric data are available for many AGNs. Support for our values for the IR lags for Ark~120 and Mrk~509 comes from them being in good agreement with the luminosity--IR delay relation (Oknyanskij 1999; Okyanskij \& Horne 2001). As can be seen from Table \ref{tab1}, for most of the objects, time lags at different IR bands are similar. Only for one object, GQ Comae, do we see difference at about a factor of three between lags in the $L$ and $K$ bands. A similar difference was seen for NGC~4151 but only during the long, very high state which was the only one during last 110 years (Oknyanskij et al. 2013). The probability of observing an object exactly in its highest state is not large. Perhaps this is the reason why we see similar IR lags at different wavelengths for most of objects. The similar IR lags at the near and mid-IR were recently predicted in the model of a compact IR region as a part of dust torus by H\"onig \& Kishimoto (2011). They considered the case in which the dust is located much farther then the sublimation distance. For the more realistic case of an IR region that is not too compact, the predicted values of the  peak of the cross-correlation functions are too low in comparison to observed ones. Some new observations show that hot dust clouds radiating in the mid-IR are located more in the polar regions (H\"onig et al. 2013) than in the torus. Those results were explained by optically-thin, dusty wind which is launched from the hottest and inner region of an optically-thick dusty disk. The beginning of this outflow with the hottest dust can be connected also with near-IR radiation. So we considered an alternative model where the variable near- and mid-IR radiation arises from the near side of a hollow, bi-conical outflow of dust clouds.\\[2mm]
\begin{figure}[h]
\resizebox{\hsize}{!}
{\includegraphics{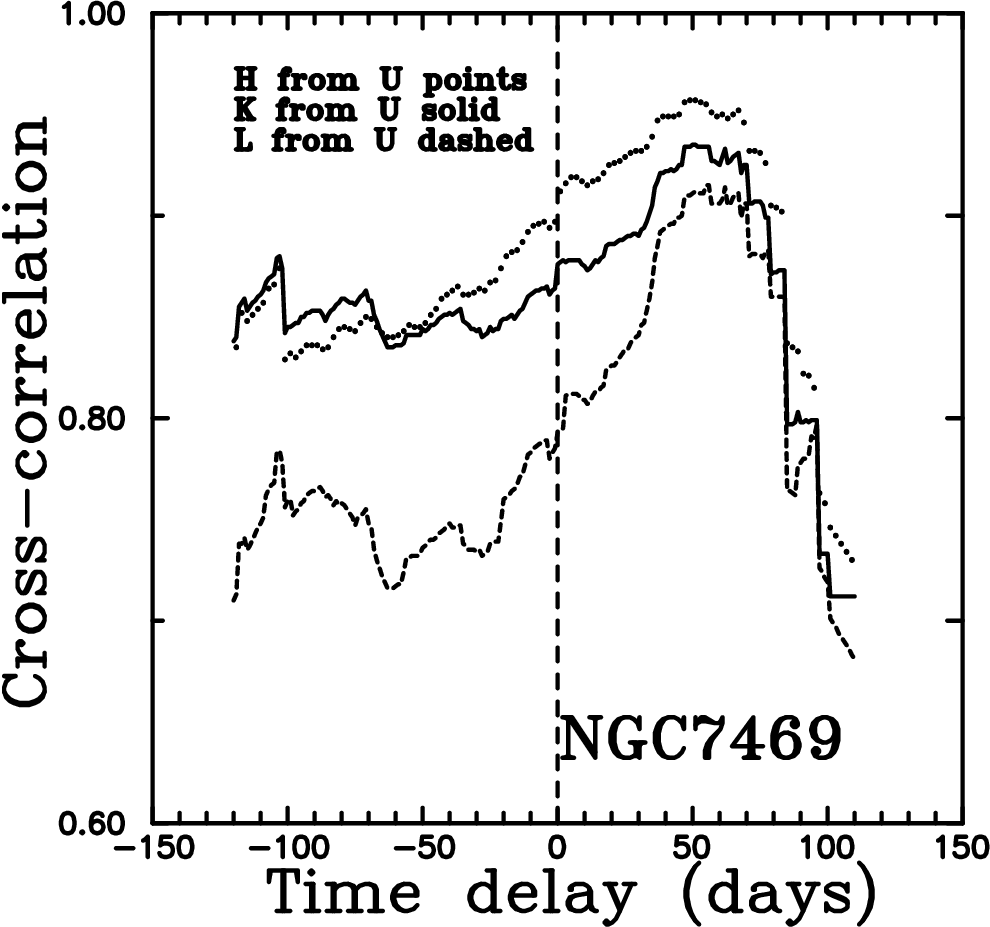}}
\caption{Cross-correlation functions calculated with the MCCF method for NGC~7469 using IR  (Glass 1998) and optical   (Lyuty  1995)  observations  during  2008-2013.  See also Oknyansky (1999) and Oknyanskij \& Horne (2001) where the IR lags for the object were found first time.}
\label{fig1}
\end{figure}

\begin{table*}[h]
\small
\caption{Infrared lags in different bands, previously published and measured in this work.}
\begin{tabular}{llllcc}
\hline
Object&Time delay,&Bands&Time&References&References to data\\
& days (1 from 2) & 1(2) & Interval &  to results &(for new results)
      
\\
\hline
NGC4151& $18\pm 6$ &K(U)&1969-1980&Oknyanskij (1993)&\\
  & $24\pm 6$ & L(U)& & Oknyanskij \& Horne(2001)&  \\
  & $35\pm 8$ & K(U)& 1985-1998& Oknyanskij et al. (1999)&  \\
  & $8\pm 4$ & H(U)& & &  \\
  & $97\pm 10$ & L(U)& & &  \\
  & $104\pm 10$ & K(U)&1998-2003 & Oknyanskij et al (2006) &  \\
  & $94\pm 10$ & H(U)& & &  \\
  & $105\pm 10$ & L(U)& & &  \\
  & $41\pm 5$ & KH(U)&2003-2006 & Oknyanskij et al.  (2008) &  \\
  & $105\pm 5$ & L(U)& & &  \\
  & $94\pm 10$ & L(J)& & &  \\
  & $37\pm 5$ & K(U)&2003-2007 & &  \\
  & $40\pm 6$ & JHKL(B)&2008-2013 & Oknyansky et al. (2014a,b) &  \\
NGC 6418  & $37\pm 2$ & 3.6$\mu$m(B) &2011-2013 & Vazquez  et al.(2015) &  \\
  & $47\pm 3$ & 4.5$\mu$m(B) & &  & \\
  & $40\pm 5$ & 3.6$\mu$m(B) & & This work & Vazquez  et al.(2015) \\
  & $50\pm 5$ & 4.5$\mu$m(B)& & &  \\
NGC 7469  & $52\pm 15$ & K(U)&1984-1996 & Oknyanskij \& Horne(2001) & Glass(1998), Luyty(1995) \\
  & $60\pm 10$ & L(U)& & &  \\
  & $50\pm 10$ & H(U)& & This work &  \\ 
  & $88$ & K(V)&2001 &Suganuma et al.(2006) &  \\
  & $54$ & K(V)&2002 & &  \\
  & $87$ & K(V)&2001-2002 & &  \\
  & $70\pm 5$ &H(V)&2001 &This work & Suganuma et al.(2006) \\
  & $88\pm 5$ &K(V)& & &  \\
  & $48\pm 5$ &H(V)&2002 & &  \\
  & $53\pm 5$ &K(V)& & &  \\
NGC 5548  & $47$ &K(V)&2001-2003 & Suganuma et al.(2006) &  \\
  & $48$ &K(V)&2001 & & Suganuma et al.(2006) \\
  & $48 \pm 4$ &K(V)&2001-2003 & This work &  \\
  & $37 \pm 6$ &H(V)& & &  \\
  & $55 \pm 4$ &K(V)&2001 & &  \\
  & $52 \pm 4$ &H(V)& & &  \\
NGC 3783  & $41 \pm 8$ &J(B)&2006-2009 &Lira et al. (2011) &  \\
  & $66 \pm 6$ &H(B)& & &  \\
  & $76 \pm 14$ &K(B)& & &  \\
  & $74 \pm 5$ &J(B)& &This work &Lira et al. (2011) \\
  & $75 \pm 8$ &H(B)& & &  \\
  & $86 \pm 8$ &K(B)& & &  \\
F9& $-20 \pm 100$ &J(UV)&1982-1998 &Clavel et al.(1989) &  \\
& $385 \pm 100$ &K(UV)& & &  \\
& $250 \pm 100$ &H(UV)& & &  \\
& $410 \pm 110$ &L(UV)& & &  \\
GQ Comae& $250$ &K(UV)&1980-1989 &Sitko et al. (1993) &  \\
  &$700$ &L(UV)& & &  \\
  &$260 \pm 20$ &K(UV),K(V)& &Oknyanskij (1999) &Sitko et al. (1993)  \\
  &$760 \pm 20$ &L(V)& & &  \\
WPVS48  &$66 \pm 4$ &J(B)& 2013 &Pozo Nu{\~n}ez et al. (2014) &  \\
  &$74 \pm 5$ &K(B)& & &  \\
  &$68 \pm 4$ &K(B)& &This work &Pozo Nu{\~n}ez et al. (2014)  \\
PGC 50427  &$48 \pm 2$ &J(B)&2013 &Pozo Nu{\~n}ez et al. (2015) &  \\
  &$47 \pm 2$ &K(B)& & &  \\
M 509 &$100$ &K(U)&1985-2001 &Glass (2004) &  \\
  &$60$ &L(U)& & & \\
  &$104 \pm 20$ &J(B)& &This work &Glass (2004)  \\
  &$139 \pm 20$ &HK(B)& & & Chuvaev et al.(1997)\\
  &$153 \pm 20$ &L(B)& & &\\
MCG-6-30-15 &$11 \pm 4$ &J(B)&2006-2011 &Lira et al.(2015) &\\
  &$17 \pm 4$ &H(B)& & &\\
  &$19 \pm 4$ &K(B)& & &\\
Ark 120  &$315$ &U(L)&1986-2001 &Glass (2004) &  \\
  &$95 \pm 30$ &JHKL(B)& &This work &Glass (2004)  \\
  & && & &Doroshenko \& Lyuty(1999) \\
\hline
\end{tabular}
\label{tab1}
\end{table*}

{\bf 3. A hollow, bi-conical, outflow model}\\[1mm]

\begin{figure}[h]
\resizebox{\hsize}{!}
{\includegraphics{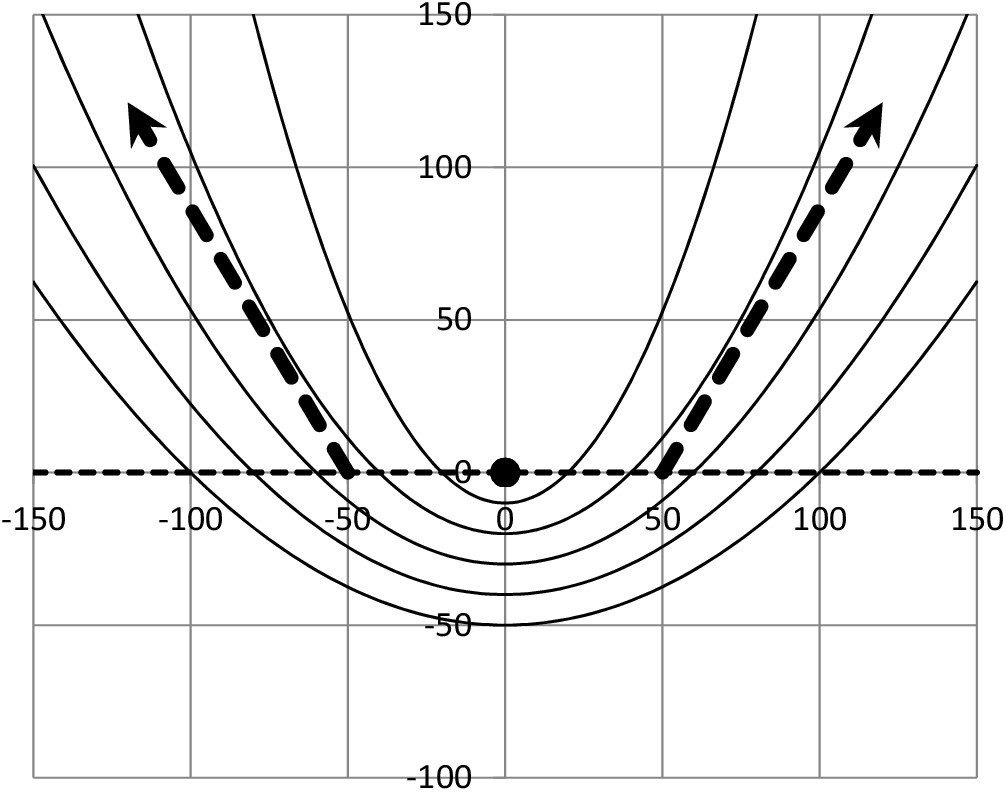}}
\caption{Cross section (in light days) of the inner wall of the near side of a hollow, bi-conical outflow (thick dashed arrows).  Iso-delay paraboloids are shown at 20 day intervals.  The horizontal dashed line shows the location of the optically-thick mid plane which obscures our view of the other side of the bi-cone (not shown).}
\label{fig2}
\end{figure}

\begin{figure}[h]
\resizebox{\hsize}{!}
{\includegraphics{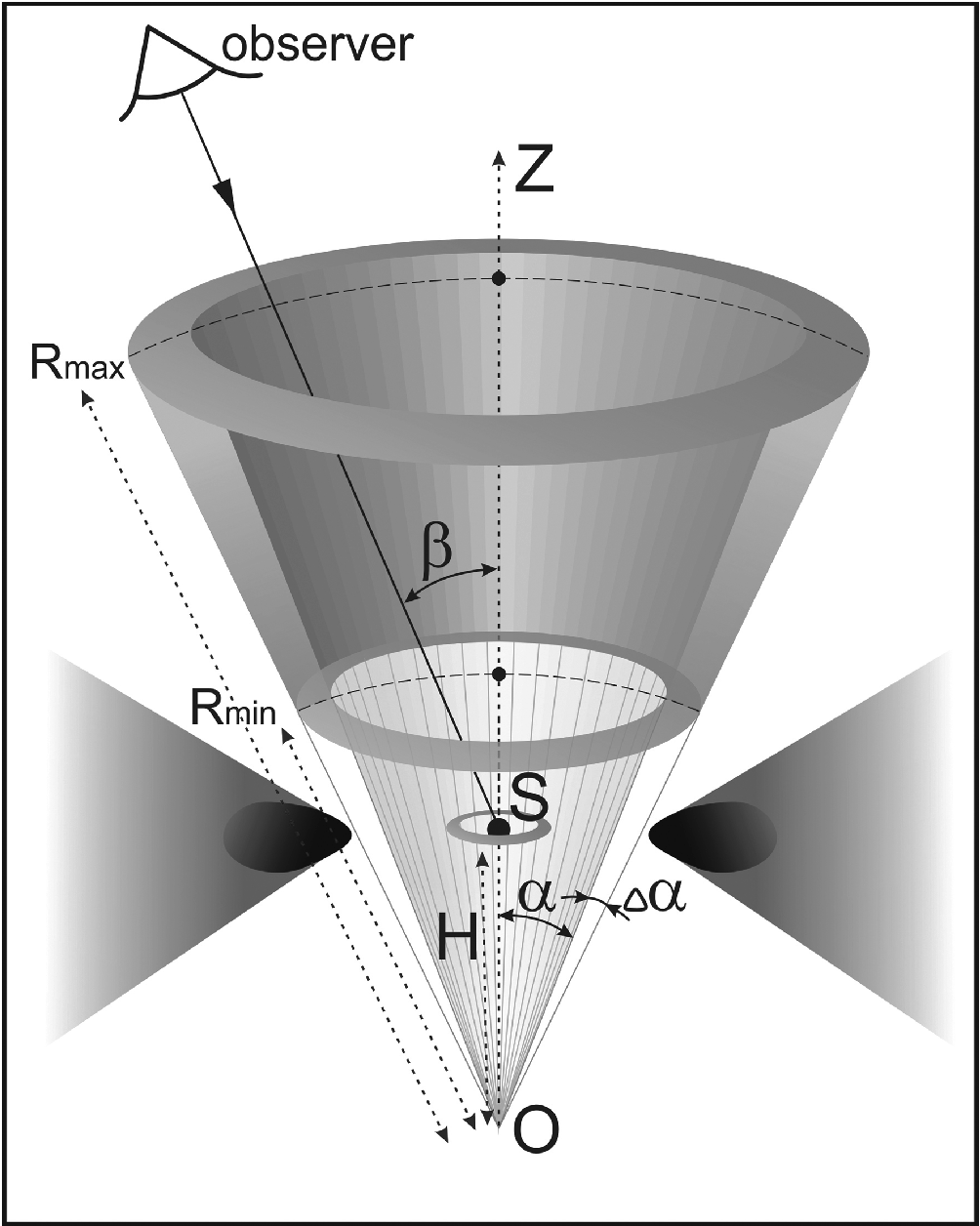}}
\caption{Geometry of the near side of the proposed hollow bi-conical dust distribution.}
\label{fig3}
\end{figure}

If, as it is often depicted, the dust is in a flattened distribution in a plane approximately perpendicular to the line of sight, the lag, $\tau$, gives the distance, $c\tau$, of the emitting region from the central source. Therefore, if the IR lags are similar, as seems here usually to be the case, dust of different temperatures would be at approximately the same radii. An alternative possibility, that does not require dust of different temperatures to be at similar radii, is that the hot dust is the inner surface of a hollow conical outflow. As it is shown in Fig. \ref{fig2}, the part of the cone on the nearside of the AGN has its surface approximately tangential to the iso-delay surfaces for response to variations of the continuum when viewed from near face-on (as is usually the case for type-1 AGNs).
The cone on the far side gives a much more spread out response and, since the material in the equatorial plane (what will be come the accretion disc) is optically thick, the response of the far side of the cone is probably not seen.  We show the inner wall of the near side of the hollow bi-cone in Fig.\ref{fig3}.

IR and optical light curves of an object are connected
through a response function $\Psi(\tau)$ that depends
on geometry and physical properties of the emitting
medium:
\begin{equation}
F_{IR}(t) \sim const + \int_{-\infty}^{\infty} \Psi (\tau) F_{opt}(\tau - t) d \tau
\label{respfun}
\end{equation}

$\Psi(\tau)$ can be explained as a response of the medium to a central source's UV impulse in the form of the $\delta$-function. We assume that the optical and UV variability occurs almost simultaneously. We can use $\Psi(\tau)$ from a model or we can try to estimate it from the real data using some methods for solving ill-posed inverse problems (see the next paragraph). Then we can investigate the structure and physical properties of the emitting medium through comparison of response functions, obtained from observational data, with ones, predicted by different models.

We have calculated $\Psi(\tau)$ directly for our model {\it via} Monte-Carlo simulations with 10000 dust cloudlets distributed randomly within the hollow bi-conical space given by parameters $H$, $R_{min}$, $R_{max}$, $\beta$, $\alpha$, $\Delta\alpha$ (see Fig. \ref{fig3}). We also take into account the dependence of the UV flux on distance to the central source $S$ and the anisotropy of the UV radiation. To get the $\Psi(\tau)$, we consider a short $\delta$-function UV pulse at $t=0$ and treat each cloud as a point-like object which re-radiates in the IR only when it receives the UV pulse at a time lag $\tau$. The lag is simply light travel time for the fixed distance from $S$ to the cloud.

\begin{figure}[h]
\resizebox{\hsize}{!}
{\includegraphics{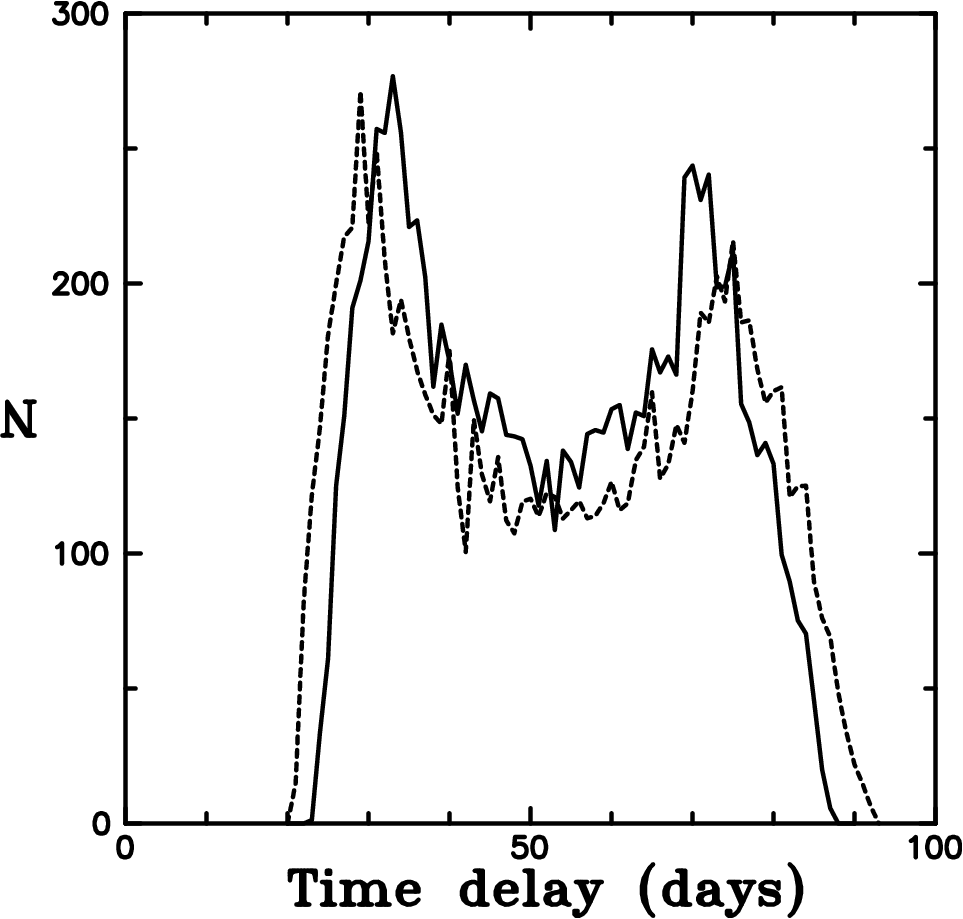}}
\caption{Monte Carlo simulations (using 10,000 clouds) of response functions for thin conical shells. $H = 90$ ld,    $\beta = 20^\circ$, $\alpha = 30^\circ$, $\Delta \alpha = 5^\circ$, $F_{UV} \sim 1 /R^2$, axial anisotropy of the UV radiation is given by $F_{UV} \sim [cos(\gamma) (1 + 2 cos(\gamma))/3]^{1/2}$ (Netzer 2015). Solid line: $R_{min} = 120$ ld, $R_{max} = 140$ ld; dashed  line: $R_{min} = 140$ ld, $R_{max} = 160$ ld.  $N$ is the number of clouds responding per day.}
\label{fig4}
\end{figure}

We obtain two response functions for two different  regions: one close and one further away from $S$ (see Fig. \ref{fig4}).
 \\[2mm]

{\bf 3. Reconstruction of response functions from real data}\\[1mm]

Unfortunately, we cannot obtain $\Psi(\tau)$ directly from (\ref{respfun}), because the observational data are unevenly spaced and have long gaps. So we propose a  new approach for recovery of the response function that combines the cross-correlation analysis of observational unevenly spaced time series and Tikhonov regularization.

As can be easily shown, the cross-correlation function $CCF_{F_{IR}F_{opt}}$  is a convolution of the auto-correlation function $ACF_{F_{opt}}$  with $\Psi(\tau)$:
\begin{equation}
CCF_{F_{IR}F_{opt} }\sim \int_{-\infty}^{\infty} \Psi (\tau) ACF_{F_{opt}}(\tau - t) d \tau
\label{ccf}
\end{equation}

There are several methods of cross-correlation analysis which allow obtaining cross-correlation function for unevenly spaced time series. We use the MCCF method (see details and references at Oknyansky et al. 1999, 2014a). The advantage of the MCCF method is that it imposes a limit on the interpolation which is necessary for unevenly spaced data.  For our purposes we used only the central part of the MCCF peak with the length 80 days ($\pm 40$ days from the peak delay). 

The equation (\ref{ccf}) is a Fredholm integral equation of the first kind with the convolution kernel. This is an ill-posed problem. Such problems can be solved using the Tikhonov regularization approach (Tikhonov et al. 1995) that allows one to find approximate solution, taking into account {\it a priori} information about the function sought, e.g., the smoothness of the solution or its closeness to some model (Koptelova et al. 2005).

Assuming that $\Psi(\tau)$ is a square integrable function, we compose the Tikhonov smoothing function:
\begin{equation}
M^\alpha[\Psi] = \Vert CCF_{F_{IR}F_{opt} } - ACF_{F_{opt}} \ast \Psi \Vert_{L_2}^2 + \alpha \Vert \Psi \Vert _{L_2}^2
\label{tikhf}
\end{equation}

Provided that the regularization parameter $\alpha$ is chosen according to the discrepancy principle, the solution $\Psi^\alpha$ of the minimization problem for $M^\alpha[\Psi]$ can be considered as an approximate solution of (\ref{ccf}).

Results of the reconstruction of the response
function for the Seyfert galaxy NGC 7469 with the
suggested approach based on observational $K$ and
$V$ light curves in 2002 (Suganuma et al. 2006) are
presented in Fig. 5. The reconstructed response
function has two peaks. Two peaks are also
observed in the theoretical response function in
Fig. 4. Perhaps this is a characteristic feature
of our proposed  model. It will be of interest to
investigate this in other objects.
\\[2mm]

\begin{figure}[h!]
\resizebox{\hsize}{!}
              { \includegraphics{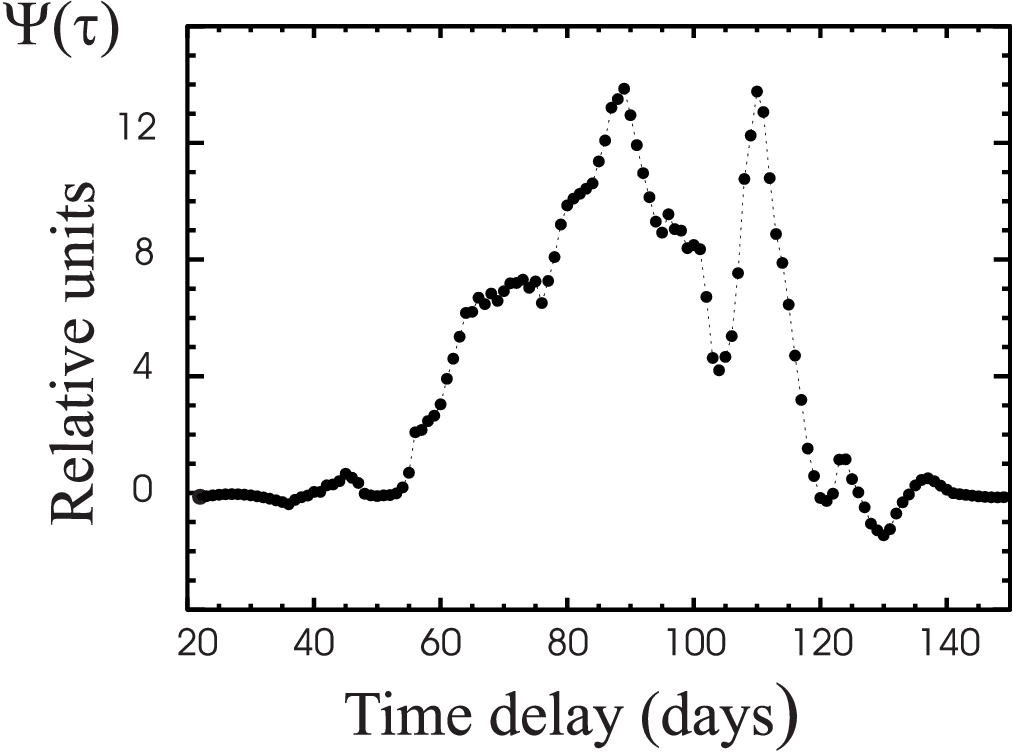}}
\resizebox{\hsize}{!}
             { \includegraphics{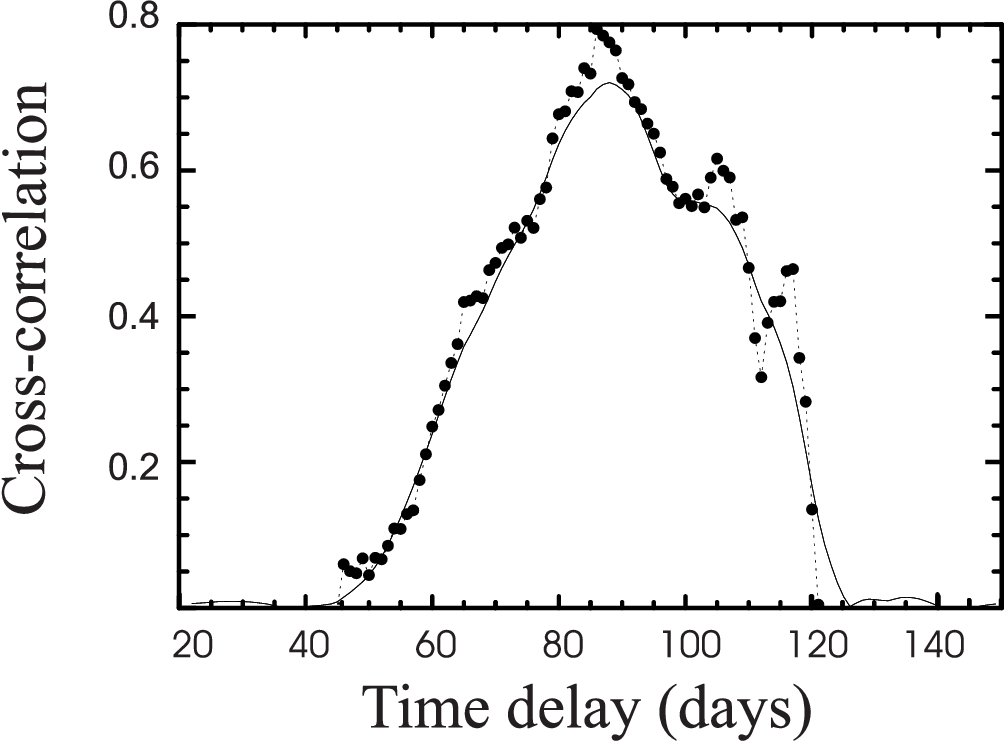}}

\caption{Upper: the reconstructed response function $\Psi(\tau)$ for the Seyfert galaxy NGC 7469. Lower: $CCF_{F_{K}F_{V}}(t)$, calculated from K and V band lightcurves (dotted line), and $ACF_{F_{V}} \ast \Psi$ for the reconstructed response function (solid line).}
\label{tikh}
\end{figure}

{\bf 4.	Cosmological applications}\\[1mm]

The ability to measure cosmological constants using the time delays of the near-infrared variability relative to the optical one was first mentioned by Kobayashi et al. (1998) and independently by Oknyanskij (1999, 2002). At the same time it was first shown (Oknyanskij 1999, 2002; Oknyanskij \& Horne 2001) that for a number of AGNs the time delay of variability (in the $K$ band) depends on the UV luminosity as $\tau \propto  L_{UV}^{1/2}$ in accordance with the theoretical result of Barvainis (1987, 1992). In the recent publications Yoshii et al. (2014) and H\"onig (2014) considered in detail the possibility of estimating cosmological parameters on the basis of measuring the dust sublimation radii in AGNs. Of particular interest is the ability to measure distances to  objects with large redshifts, $z$. For large $z$ it is necessary to take into account the possible dependence of the delays on wavelength in the IR range and to make any appropriate correction. The equation for such a correction was derived by Oknyansky \& Horne (2001) on the basis of the theory (Barvanis 1987) and observational data available at that time. However, as we have found now, the IR delays do {\em not} depend significantly on wavelength for most  AGNs.  The significant difference in IR lags with wavelengths can be just a temporary, rare situation in particular objects, as happened in the case of NGC 4151 just when it was in its very high state.  Therefore for most of cases for a wide range of $z$ the IR lags  can be easy corrected to the rest frame just by allowing for the time dilation  (i.e., dividing by $(1 + z)$).
\\[2mm]

{\bf 5. Conclusions}\\[1mm]

We find that most AGNs show at best only a small increase of lags in the $J$, $H$, $K$, and $L$ bands.  Just one object, GQ Comae (which is a high-luminosity AGN), probably has big differences in the IR lags (lag for $L$ is about 3 times bigger than for $K$). A similar difference in IR lags was observed temporarily in NGC~4151 when it was in a very high state. Probably this big difference in IR lags is at epochs when the dust is intensively sublimated. During more normal times when the dust region is located much farther out than the sublimation distance, the time lags are about the same.

Our model with hollow bi-conical outflow of the dust clouds does not contradict to the existence of the dust torus. Moreover, the torus and accretion disc in the model are needed to block the IR light from farther part of the conical outflow. Also the torus can be a reservoir for the dust in the outflow.

In the Table \ref{tab1}, a significant number  of the objects (5 out of 11) with similar IR lags have been observed to change their Seyfert type. This could just be the result of these objects having been studied particularly well. At the same time our model can explain cases of changing Seyfert type. In the very low luminosity states of some objects, when they look like Seyfert 2s, dust creation can commence in the gas clouds in the hollow of the conical outflow. So the thickness of the conical shell with dust clouds will be bigger and some of the dust clouds can be on the line of sight and block the broad line region from view and the object will be classified as a type 2 (see the same idea for the torus model in Mason 2015).

Our model can explain not only the similar IR lags at different wavelengths but also the high correlation values for IR and optical variations. Furthermore, the model can better explain the observed luminosity in IR since the IR-emitting region (i.e., the dust cone) has a higher covering factor and hence is getting more UV radiation than in the case of the flat compact IR-emitting region.

Finally, we note that the relative wavelength independence of time lags in the IR simplifies the use of these time delays to estimate luminosities and hence estimate cosmological parameters.
\\[2mm]

{\it Acknowledgements.} We are thankful to Sebastian H\"onig for useful comments, to Boris Artamonov for fruitful discussions, and to graphic designer Natalia Sinugina for producing Figure \ref{fig3}. This work has been supported by the Russian Foundation for Basic Research through grant no. 14-02-01274.
\\[3mm]
{\bf References\\[2mm]}
Barvainis, R.: 1987, \textit{Ap.J.}, \textbf{320}, 537.\\
Barvainis, R.: 1992 \textit{Ap.J.}, \textbf{400}, 502.\\
Clavel, J. et al.: 1989, \textit{Ap.J.}, \textbf{337}, 236. \\
Chuvaev, K.K. et al.: 1997, \textit{Ast.~Lett.}, \textbf{23}, 355.\\
Doroshenko, V.G.; Lyuty, V.M.: 1999, \textit{Ast.~Lett.}, \textbf{25},771.\\
Glass, I.: 1998, \textit{MNRAS}, \textbf{297}, 18.\\
Glass, I.: 2004, \textit{MNRAS}, \textbf{350}, 1049.\\
H\"onig S.F.; Kishimoto M.: 2011.  \textit{A\&A}, \textbf{524}, A121.\\
H\"onig S.F. et al.: 2013,  \textit{Ap.J.}, \textbf{771}, 87.\\
H\"onig S.F.: 2014,  \textit{Ap.J.~Lett}, \textbf{774}, L4.\\
Kobayashi  Y. et. al.: 1998, \textit{Proc.~SPIE},  \textbf{3352}, 120.
Koptelova E. et al.: 2005, \textit{MNRAS}, \textbf{356}, 323.\\
Lira, P. et al.: 2011, \textit{MNRAS}, \textbf{415}, 1290.\\
Lira, P. et al.: 2015, \textit{MNRAS}, \textbf{454}, 368.\\
Lyutyi, V.M. et al.: 1995,  \textit{Ast.~Lett.}, \textbf{21}, 581.\\
Mason, R.E.: 2015, \textit{P\&SS}, \textbf{116}, 97.\\
Netzer, H.: 2015, \textit{ARA\&A}, \textbf{53}, 365.\\
Oknyanskij, V.L: 1993 \textit{Ast.~Lett.}, \textbf{19}, 416.\\
Oknyanskij, V.L.: 1999, \textit{Odessa Ast.~Pub.}, \textbf{12}, 99. \\
Oknyanskij, V.L. et al.: 1999, \textit{Ast.~Lett.}, \textbf{25}, 483.\\
Oknyanskij, V.L.; Horne, K.: 2001, \textit{ASP Conf.~Proc.}, \textbf{224}, 149.\\
Oknyanskij, V.L. et al.:2006, \textit{ASP Conf.~Ser.}, \textbf{360}, 75.\\
Oknyanskij, V.L. et al.:2008, \textit{Odessa Ast.~Pub.}, \textbf{21}, 79.\\
Oknyanskij, V.L.: 2002, \textit{ASP Conf.~Proc.}, \textbf{282}, 330.\\
Oknyanskij, V.L. et al.: 2013, \textit{Odessa Ast.~Pub.}, \textbf{26}, 212.\\
Oknyansky, V.L. et al.: 2014a, \textit{Ast.~Lett.}, \textbf{40}, 527.\\
Oknyansky, V.L. et al.: 2014b,\textit{Odessa Ast.~Pub.},\textbf{27},47.\\
Pozo Nu\~nez et al.: 2014, \textit{A\&Ap}, \textbf{561}, L8.\\
Pozo Nu\~nez et al.: 2015, \textit{A\&Ap}, \textbf{576}, 73.\\
Sitko, M.L. et al.: 1993,  \textit{Ap.J.}, \textbf{409}, 139.\\
Suganuma, M. et al.: 2006, \textit{Ap.J.}, \textit{639}, 46.\\
Tikhonov A.N., Goncharsky A.V., Stepanov V.V., Yagola A.G.: 1995, \textit{Numerical Methods for the Solution of Ill-Posed Problems}. Kluwer Academic Press, Dordrecht.\\
Vazquez.  B. et al.: 2015, \textit{Ap.J.}, \textbf{801}, 127.\\
Yoshii Y. et al.: 2014, \textit{Ap.J.~Lett}., \textbf{784}, L11.
\vfill
%

\end{document}